\documentstyle[preprint,aps]{revtex}
\begin{document}
\draft
\title{A New Exponent Characterizing the Effect of Evaporation on
Imbibition Experiments}

\author{L. A. N. Amaral,$^1$ A.-L. Barab\'asi,$^1$ S. V. Buldyrev,$^1$
S. Havlin,$^{1,2}$ and H. E. Stanley$^1$}

\address{$^1$Center for Polymer Studies and Dept. of Physics,
 Boston University, Boston, MA 02215 USA \\ $^2$Dept. of Physics,
 Bar-Ilan University, Ramat Gan, Israel}

\date{\today}

\maketitle

\begin{abstract}

We report imbibition experiments investigating the effect of evaporation
on the interface roughness and mean interface height.  We observe a new
exponent characterizing the scaling of the saturated surface width.
Further, we argue that evaporation can be usefully modeled by
introducing a gradient in the strength of the disorder, in analogy with
the gradient percolation model of Sapoval {\it et~al.}.  By
incorporating this gradient we predict a new critical exponent and a
novel scaling relation for the interface width.  Both the exponent value
and the form of the scaling agree with the experimental results.

\end{abstract}


\narrowtext

Recently the growth of rough interfaces has witnessed a veritable
explosion of theoretical and experimental results, partly fueled by the
broad interdisciplinary aspects of the subject \cite{VFK}.  Much
attention has focused on measuring the roughness exponent $\alpha$,
defined by the power law dependence upon the observation length scale
$\ell$ of the width $w(\ell)$. Simulations on discrete models provide
exponents in agreement with the predictions of phenomenological
continuum theories \cite{KPZ86}.  However, experimental studies find
exponents significantly larger than the predictions of
theory ---for example, for dimension $d=(1+1)$, theory predicts
$\alpha=1/2$ but experiments show $\alpha \simeq 0.63-0.8$
\cite{RHV,Buldyrev}.
Moreover, experimental studies frequently detect a crossover in
$w(\ell)$ to a different behavior above some characteristic length
$\ell_{\times}$.
It is currently believed that the anomalously large values of the
exponent $\alpha$ is due to quenched pinning disorder
\cite{VFK,Buldyrev,Tang92a}, however, a completely satisfactory
explanation of the experimentally determined crossover length has not
yet been found.

Here we present imbibition experiments that probe the effect on the
growth process of the evaporation rate and suspension concentration.  We
find that the scaling of the interface width changes with the
evaporation rate and is characterized by a new exponent $\gamma$.  We
also present a model, inspired by the model of Ref.
\cite{Buldyrev}, that predicts the experimentally-observed value of
the new exponent characterizing the crossover effect.

  The key ingredient in the model is to allow for a gradient in the
density of the pinning cells, which results in the stopping of the
interface.  Moreover, the detailed investigation of the scaling
properties of the model provides us with a new scaling law and new
critical exponents.  Using this scaling law, we find good data collapse
and scaling exponents that agree with the values determined analytically
and numerically.

In our experiments, paper --- the ``disordered medium'' --- is dipped
into a reservoir filled with a colored suspension (coffee, ink) and the
propagating wetting front is observed.
The wetting front reaches a critical height, $h_c$, above the level of
the liquid, and stops propagating when the evaporation of the liquid
induces the pinning of the interface by the inhomogeneities of
the paper.  We digitize the rough boundary between colored and uncolored
areas  and measure a roughness exponent $\alpha \simeq 0.63$
\cite{Buldyrev}.

Although the experiments are straightforward, their explanation
in physical terms is less so. At microscopic length scales, paper is an
extremely disordered substance, formed by long fibers that are randomly
distributed and have random connections among one another.  The wetting
fluid propagates in these fibers mainly due to capillary forces, but the
random nature of the fiber network and the particles in the suspension
provide constant obstacles for the fluid flow \cite{expl1}.
As we depart from the water source, evaporation is
constantly decreasing the fluid pressure, making it more and more difficult
for the fluid to overcome these microscopic ``obstacles''.  At the critical
height, the fluid pressure balances the effect of the pinning
obstacles and the fluid stops propagating \cite{expl2}.

We anticipate, on physical grounds, that the smaller the evaporation,
the larger the critical height will be. To check this intuition we
repeated our experiments in environments with different rates of
evaporation. As we decreased the evaporation rate, the height reached by
the interface increased sharply.
These effects can be observed in the pinned interfaces obtained
experimentally (Fig. \ref{fig1}(a)).

While the above results can be understood from microscopic
considerations, the effect of the evaporation or of the suspension
density on the scaling of the interface roughness is nontrivial.

To understand these results, we propose a model to describe the formation
of the pinned interface in a disordered medium.
In $(1+1)$ dimension, we model the pinning obstacles by blocking, in a
lattice of horizontal size $L$, a fraction $p(h)$ of the cells in each
horizontal row, where $h$ is the height from the bottom of the lattice.
We start, at $t=0$, from a horizontal line of wet cells at the bottom
edge of the lattice.
At time $t+1$ we wet all unblocked cells which are {\it nearest
neighbors to the wet region\/} at time $t$. We also apply the rule that
every cell, {\it blocked or not\/}, bellow a new wet cell becomes wet as
well (Fig. \ref{fig2}). The motivation for this rule is the experimental
observation that the wet region is, at least at macroscopic length
scales, nearly free of dry islands.

If $p(h)=p_o$, this model generates an interface which propagates with a
constant ($p$-dependent) velocity if $p_o < p_c \simeq 0.47$, and becomes
pinned by a {\it directed percolation cluster} that spans the system at
$p_o \ge p_c$ \cite{Buldyrev,Tang92a}.
However, although the actual disorder in the paper {\it is not height
dependent}, its {\it effect\/} in pinning the propagation of the fluid
{\it is increasing with height}, due to the decrease in the fluid pressure.
The most physical assumption is an exponential decrease of the fluid
pressure or, equivalently, of the driving force. This will lead to
an ``effective'' increase in the density of pinning obstacles
\cite{expl2} as we depart from the reservoir, i.e., $p \equiv p(h)$.
Hence
\begin{equation}
\ p(h) - p_o \propto 1 - e^{-h/h_o}.
\label{ppp}
\end{equation}
If $h \ll h_o$ and $p_c - p_o \ll p_c$, we can write
\begin{equation}
\ p(h) - p_o \propto h_o^{-1} h \propto (\nabla p)h.
\label{p}
\end{equation}
Hence, in this limit, we find a constant {\it non-zero gradient} in the
density of pinning obstacles.

The presence of the gradient $\nabla p$ changes the width of the pinned
interface (Fig. \ref{fig1}(b)) and its scaling form (Fig. \ref{fig3}).
Our simulations show that for observation scales $\ell$ much smaller
than some characteristic crossover length $\ell_\times$, the saturated
width behaves as $w \sim \ell^\alpha$, but for $\ell \gg \ell_\times$,
the width saturates at a value $w_{\rm sat}$ that depends upon the gradient
as
\begin{equation}
\ w_{\rm sat} \sim (\nabla p)^{-\gamma}.
\label{w1}
\end{equation}
This behavior can be expressed by a scaling law of the form
\begin{mathletters}
\begin{equation}
\ w(\ell, \nabla p) \sim \ell^\alpha f({\ell}/\ell_\times)
\end{equation}
\begin{equation}
\ \ell_\times \sim {(\nabla p)^{-{\gamma}/{\alpha}}}.
\end{equation}
\label{w2}
\end{mathletters}
The scaling function $f(u)$ satisfies $f(u \ll 1) \sim const$ and
$f(u\gg 1) \sim u^{- \alpha}$.  Our simulations (see Fig. \ref{fig3})
for a system of size $L=16384$ yield the exponents
\begin{equation}
\ \alpha_{\rm sim} = 0.63 \pm 0.02, \quad \gamma_{\rm sim} = 0.52 \pm 0.02.
\label{ex1}
\end{equation}
We remark that the validity of the scaling law (\ref{w2}) and the values
of the exponents do not depend on the exact form of $p(h)$ but only on the
value of $\nabla p(h)$ at $h_c$ \cite{Sapoval}.

The value of $\alpha$ can be understood from the mapping to directed
percolation, since the conditions for a complete pinning of the
interface do not change from the models of Refs.
\cite{Buldyrev,Tang92a}.
In directed percolation, the size of a cluster is characterized by a
longitudinal correlation length $\xi_{\parallel}$ and a transverse
correlation length $\xi_{\perp}$ that, near $p_c$, behave as
\begin{equation}
\ \xi_{\parallel} \sim |p_c-p|^{-\nu_{\parallel}}, \quad \xi_{\perp} \sim
|p_c-p|^{-\nu_{\perp}}.
\label{xi}
\end{equation}
The roughness exponent is related to the exponents of directed
percolation as \cite{Buldyrev,Tang92a}
\begin{equation}
\ \alpha = \nu_{\perp}/\nu_{\parallel}.
\label{alp}
\end{equation}
Using the known values of $\nu_{\perp}$ and $\nu_{\parallel}$
\cite{EE} in relation (\ref{alp}) we predict $\alpha = 0.633 \pm
0.001$, in agreement with our simulation result (\ref{ex1}).

The exponent $\gamma$ can be related to $\nu_{\perp}$ theoretically.
A point of the interface, at distance $w_{\rm sat}$ of the critical height,
is pinned by a directed percolation cluster if the transverse size of
that cluster is of order $\xi_{\perp}(p)$. At that point we have
$p=p(h_c \pm w_{\rm sat}) \approx p_c \pm w_{\rm sat} \cdot \nabla p$.
Therefore, using Eq. (\ref{xi}) we find \cite{Sapoval,Hansen}
\[
\ w_{\rm sat} \sim \xi_{\perp}(p) \sim |p_c - (p_c \pm w_{\rm sat} \cdot
\nabla p)|^{-\nu_{\perp}},
\]
\begin{equation}
\ w_{\rm sat} \sim |w_{\rm sat} \cdot \nabla p|^{-\nu_{\perp}}.
\label{gama3}
\end{equation}
 From Eqs. (\ref{w1}) and (\ref{gama3}) follows
\begin{equation}
\ \gamma = {\nu_{\perp}}/({1 + \nu_{\perp}}).
\label{gama2}
\end{equation}
Since $\nu_{\perp}$ is known accurately \cite{EE}, Eq. (\ref{gama2})
predicts $\gamma = 0.523 \pm 0.001$, in excellent agreement with our
simulation result (\ref{ex1}).

Without the gradient, the interface has critical behavior only if we tune
$p$ to $p_c$. However, with the gradient the interface always stops at
the critical height $h_c$.  This critical height can be calculated from
the condition $p(h_c)=p_c$. Thus from (\ref{p}) we obtain
\begin{equation}
\ h_c \sim (\nabla p)^{-1},
\label{h_c}
\end{equation}
i.e. the height reached by the wetting fluid is inversely proportional
to the gradient in the disorder.

The experimental data presented in Fig. \ref{fig4}(a)
remarkably resemble the data obtained for the model. However, without
knowing the actual value of the gradient in the experiments, it is not
possible to check the validity of the scaling law (\ref{w2})
experimentally.  Nonetheless, measuring the critical height in the
experiments and using Eq. (\ref{h_c}), we are able to estimate $\nabla
p$, the gradient in the ``effective disorder'', for the experiments, up
to a multiplicative constant.  Using these experimentally determined
values of $\nabla p$, we rescale the results obtained for the width
according to the scaling law (\ref{w2}).
In Fig. \ref{fig4}(b) we show this rescaling, where we used
\begin{equation}
\ \alpha_{\rm exp}=0.65 \pm 0.05 , \quad \gamma_{\rm exp}=0.49 \pm 0.05,
\label{exexp}
\end{equation}
The experimental values of both exponents agree well with the results
obtained from the simulations (Table \ref{tab1}) and with the
theoretical predictions based in known results from directed percolation.

The generalization of the model to $d=(2+1)$ is straightforward, and in
this case the pinned interface can be mapped to directed surfaces
\cite{Buldyrev}, a percolation problem that has not been thoroughly
investigated.  We simulated the model for a $512 \times 512$ system;
the critical exponents that give the best data collapse are
\begin{equation}
\ \alpha_{\rm sim} = 0.43 \pm 0.04 , \quad \gamma_{\rm sim} = 0.32 \pm 0.02.
\label{ex2}
\end{equation}
 From these results, we calculate the exponents characterizing the
transverse and longitudinal correlation lengths for the directed
surfaces problem, obtaining
\begin{equation}
\ \nu_{\perp} = 0.47 \pm 0.04, \quad \nu_{\parallel} = 1.1 \pm 0.1.
\label{ex3}
\end{equation}

In summary, we have performed imbibition experiments to study the effect
of evaporation on interfacial phenomena. We have also
developed a model that incorporates evaporation by introducing a gradient
in the density of pinning cells \cite{expl2}. The model provides insight
into three previously unexplained aspects of imbibition experiments:
{\it (i)\/} The interface always stops growing, after some finite time.
Due to the gradient, the wetting interface only moves until it reaches a
critical concentration of pinning cells. This gradient in pinning cells
arises from the balance between the evaporation of the fluid and the
capillary forces tending to move it along the paper.
{\it (ii)\/} The final height of the interface, $h_c$, increases when the
evaporation is reduced, due to the smaller {\it effective\/} gradient in
the pinning disorder.
{\it (iii)\/} A new exponent $\gamma$ was found characterizing the
dependence on the gradient of the saturation width and the
characteristic length $\ell_\times$.
Good agreement was found between experimental, analytical and
simulation values of the exponents.

This work is part of the Ph. D. thesis of L. A. N. Amaral and A.-L.
Barab\'asi.
We thank S. Schwarzer, J. Krug and M. Ukleja for valuable contributions
and the two referees for helpful suggestions.
L. A. N. Amaral acknowledges a scholarship from Junta Nacional de
Investiga\c c\~ao Cient\'{\i}fica e Tecnol\'ogica. S. Havlin acknowledges
partial support from the Bi National US-Israel Foundation. The Center for
Polymer Studies is supported by the National Science Foundation.

\begin{figure}
\caption{Photographs of pinned interfaces in: (a) Imbibition
experiments with coffee and paper towels for {\it (i)\/} high evaporation
rate: $(\nabla p)_{exp} = 0.94 g_o$, and {\it (ii)\/} low evaporation
rate: $(\nabla p)_{exp} = 0.25 g_o$. Here $g_o$ is the undetermined
multiplicative constant discussed in the text. (b) Simulations of
the model, with $L=256$, for different values of the gradient: {\it
(i)\/} $\nabla p = 2^{-8}$ and {\it (ii)\/} $\nabla p = 2^{-10}$.
Readily apparent from these photographs is the increase in both the
final heights and widths of the interface with the decrease of the
gradient.}
\label{fig1}
\end{figure}

\begin{figure}
\caption{Example of the time evolution of the model for a very small
lattice ($L = 5$). Here, grey squares represent blocked cells and white
squares represent unblocked cells. The numbered cells are wet. The
numbers indicate at which time step the cells first become wet.
At $t=4$, we wet the cells at the left and at the right of the cell
numbered $3$. Also, in the same time step we wet the cells below those
two, regardless of the fact that they were previously blocked.
Similarly, at $t=5$, we are able to wet cells in the first column
from wet cells in the second column that were, at some earlier time,
blocked cells. The heavy line indicates the pinned interface.}
\label{fig2}
\end{figure}

\begin{figure}
\caption{The simulation results for the width $w(\ell,\nabla p)$ of the
pinned interface. (a) The widths for several values of the
gradient (averaged over 512 runs for each value of the gradient).
(b) The same simulation results, plotted in the scaling form of Eq. (3),
using the values of the exponents from Eq. (4).}
\label{fig3}
\end{figure}

\begin{figure}
\caption{The experimental results for the width  $w(\ell,\nabla p)$ of
the pinned interface.  (a) The widths for several values of the
gradient (in units of $g_o$). The values of the gradients were calculated as
described in the text; the error in these values is smaller than $10\%$.
The widths were corrected by a multiplicative factor to make them
coincide for the smallest $\ell$. (b) The same experimental
results, plotted in the scaling form of Eq. (3), using the values of the
exponents from Eq. (10).}
\label{fig4}
\end{figure}

\begin{table}
\caption{Critical probability and exponents for dimension $(1+1)$ and
$(2+1)$, calculated from the simulations of the present model.}
\begin{tabular}{ccc}
\tableline
dim.        & $(1+1)$        & $(2+1)$ \\
\tableline
$p_c$       & $0.47 \pm 0.03$ & $0.75 \pm 0.03$ \\
$\alpha$    & $0.63 \pm 0.02$ & $0.43 \pm 0.04$ \\
$\gamma$    & $0.52 \pm 0.02$ & $0.32 \pm 0.02$ \\
$\nu_\perp$ & $1.09 \pm 0.08$ & $0.47 \pm 0.04$ \\
$\nu_\|$    & $1.7 \pm 0.1$ & $1.1 \pm 0.1$ \\
\end{tabular}
\label{tab1}
\end{table}


\begin{references}
\bibitem{expl1}
In this work we refer to evaporation as the origin of the gradient.
However, the density of the suspension has a similar role in changing
the balance between capillary forces and pinning obstacles, by also
setting the value of the gradient. Indeed, our experiments suggest that
the same model accounts for not only evaporation effects, but also
suspension density effects (cf. Fig. 4).
\end{references}
\end{document}